\documentclass{tlp}

\usepackage{amsmath}
\usepackage{amssymb}
\usepackage[latin1]{inputenc}
\usepackage{url}
\usepackage{algorithm}
\usepackage{comment}
\usepackage{algpseudocode}
\usepackage{xspace}
\usepackage{booktabs}

\newcommand{\tableheadline}[1]{\textbf{#1}}
\newcommand{\fp}{\tt} % font per i programmi

\newcommand{\Fastest}{\textsc{Fastest}\xspace}
\newcommand{\setlog}{$\{log\}$\xspace}
\newcommand{\CLPPF}{\mbox{CLP(${\cal PF}$)}\xspace}
\newcommand{\CLPSET}{\mbox{CLP(${\cal SET}$)}\xspace}
\newcommand{\CLPFD}{\mbox{CLP(${\cal FD}$)}\xspace}

\newcommand{\SAT}{SAT_{\cal SET}\xspace}
\newcommand{\SATPF}{SAT_{\cal PF}\xspace}

\newcommand{\e}{\emptyset}
\newcommand{\E}{\mathsf{empty}}

\newcommand{\Int}{{\sf int}}
\newcommand{\false}{{\sf false}\xspace}

\newcommand*{\Neq}{\mathbin{\mathsf{\,neq\,}}}
\newcommand*{\In}{\mathbin{\mathsf{\,in\,}}}
\newcommand*{\Nin}{\mathbin{\mathsf{\,nin\,}}}
\newcommand*{\Subseteq}{\mathsf{subset}}

\newcommand*{\Size}{\mathsf{size}}
\newcommand*{\Integer}{\mathsf{integer}}

\newcommand{\Pfun}{\mathsf{pfun}}
\newcommand{\Dom}{\mathsf{dom}}
\newcommand{\Ran}{\mathsf{ran}}
\newcommand{\Comp}{\mathsf{comp}}
\newcommand{\Dres}{\mathsf{dres}}
\newcommand{\Rres}{\mathsf{rres}}
\newcommand{\Ndres}{\mathsf{ndres}}
\newcommand{\Nrres}{\mathsf{nrres}}
\newcommand{\Rimg}{\mathsf{rimg}}
\newcommand{\Oplus}{\mathsf{oplus}}
\newcommand{\Apply}{\mathsf{apply}}
\newcommand{\Id}{\mathsf{id}}
\newcommand{\apply}{\mathsf{apply}}

\newcommand{\set}{\mathsf{set}}
\renewcommand{\Cup}{\mathsf{un}}
\newcommand{\disj}{\mathsf{disj}}
\newcommand{\cdiff}{\mathsf{diff}}
\renewcommand{\Cap}{\mathsf{inters}}
\newcommand{\nun}{\mathsf{nunion}}

\newcommand{\ninteger}{\mathsf{ninteger}}

\renewcommand{\lor}{\,\vee}

\def\strut@op#1{\mathop{\mathstrut{#1}}\nolimits}
\def\@p#1{\mathrel{\ooalign{\hfil$\mapstochar\mkern
      5mu$\hfil\cr$#1$}}}
\def \pfun      {\@p\fun}
\let \fun       \rightarrow
\def \dom       {\mathop{\mathrm{dom}}}
\def \ran       {\mathop{\mathrm{ran}}}
\def\comp{\mathrel{\raise 0.66ex\hbox{\oalign{\hfil%
        $\scriptscriptstyle\mathsf{o}$\hfil%
        \cr\hfil$\scriptscriptstyle\mathsf{9}$\hfil}}}}
\DeclareMathSymbol{\dres}{\mathbin}{AMSa}{"43}
\DeclareMathSymbol{\rres}{\mathbin}{AMSa}{"42}
\def \ndres     {\mathbin{\rlap{\raise.05ex\hbox{$-$}}{\dres}}}
\def \nrres     {\mathbin{\rlap{\raise.05ex\hbox{$-$}}{\rres}}}

  \DeclareSymbolFontAlphabet{\bbold}{AMSb}
\def \nat       {{\bbold N}}

\def \power     {\strut@op{\bbold P}}
\def\t1{\quad}
\mathchardef\spot="320F
\mathcode`\@=\spot

\renewcommand{\iff}{\Leftrightarrow}

\newtheorem{example}{Example}
\newtheorem{definition}{Definition}
\newtheorem{theorem}{Theorem}
\newtheorem{lemma}{Lemma}

\title[Adding Partial Functions to CLP with Sets]
  {Adding Partial Functions to Constraint Logic Programming with Sets}

\author[M. Cristi\'a, G. Rossi and C. Frydman]
{MAXIMILIANO CRISTI\'A \\
CIFASIS and UNR, Rosario, Argentina \\
\email{cristia@cifasis-conicet.gov.ar}
\and GIANFRANCO ROSSI \\
Universit\`a degli Studi di Parma, Parma, Italy \\
\email{gianfranco.rossi@unipr.it}
\and CLAUDIA FRYDMAN \\
Aix Marseille Univ., CNRS, ENSAM, Univ. de Toulon, LSIS UMR 7296, France \\
\email{claudia.frydman@lsis.org}
}

\begin{document}
\maketitle

\begin{abstract}
Partial functions are common abstractions in formal specification notations such as
Z, B and Alloy. Conversely, executable programming languages usually provide little
or no support for them. In this paper we propose to add partial functions as a
primitive feature to a Constraint Logic Programming (CLP) language, namely \setlog.
Although partial functions could be programmed on top of \setlog, providing them as
first-class citizens adds valuable flexibility and generality to the form of
set-theoretic formulas that the language can safely deal with. In particular, the
paper shows how the \setlog constraint solver is naturally extended in order to
accommodate for the new primitive constraints dealing with partial functions.
Efficiency of the new version is empirically assessed by running a number of
non-trivial set-theoretical goals involving partial functions, obtained from
specifications written in Z.
\end{abstract}


\section{Introduction}\label{sec:introduction}

Given any two sets, $X$ and $Y$, a \emph{binary relation between $X$ and $Y$} is
any subset of the power set of $X \times Y$, $\power(X \times Y)$. Partial
functions are just a particular kind of binary relations, in which ordered pairs
are restricted to verify the classical notion of function---i.e. that each
element in the domain is mapped to at most one element in the range---, although
they may be undefined for some elements in the domain---i.e. they are partial. Binary
relations are in turn just sets of ordered pairs. Then, all relational operators
(such as $\dom$, $\ran$, $\comp$, etc.) can be applied to partial functions and
all set operators can be applied to both of them. Conversely, and this feature
distinguishes partial functions from binary relations, if $x$ is an element in
the domain of a partial function $f$ then $f(x)$ is defined as the element, $y$,
in the range of $f$ such that $(x,y) \in f$.

The motivation for adding partial functions to specification/programming
languages is primarily to enhance the language's expressive power. In fact,
partial functions constitute a powerful and convenient data abstraction. As an
example, the relation between the key of a table and the rest of its columns is
naturally modeled as a partial function. Partial functions are common in formal
specification notations, such as Z \cite{Spivey00}, B \cite{Abrial00} and Alloy
\cite{DBLP:conf/zum/Jackson03}, which are mainly used to specify state-based
systems (notice that, many concepts or features of these systems are best
represented as partial functions, not as total functions). Usefulness of
partial functions in executable programming languages is attested by the common
presence of library facilities, e.g. the \texttt{map} class of Java and C++,
that support at some extent the partial function abstraction. Availability of
maps, dictionaries or similar associative data structures as \emph{primitive}
components of some programming languages, such as SETL \cite{DBLP:books/daglib/0067831} or
Phyton, also attests usefulness of the partial function abstraction.

Partial functions (or maps or, more generally, binary relations) can be added
naturally also to CLP languages with sets, as observed for instance in \cite{Gervet06}.
In particular, in \cite{CristiaRossiSEFM13} we have shown how partial functions can
be encoded in the CLP language with sets \setlog (pronounced
`setlog') \cite{Dovier00}. Specifically, partial functions can be represented in \setlog as sets of
pairs, where each pair $(x,y)$ is represented as a list of two elements $[x,y]$.
Operations on partial functions can be implemented by user-defined predicates in such
a way to enforce the characteristic properties of partial functions over the
corresponding set representations.

When partial functions are completely specified this approach is satisfactory, at
least from an `operational' point of view. On the other hand, when some elements of a
partial function or (part of) the partial function itself are left
unspecified---i.e., they are represented by unbound variables---then this approach
presents major flaws. For example, the predicate {\fp ran(F,\{1\})}, which holds if
{\fp \{1\}} is the range of the partial function {\fp F}, admits infinite distinct
solutions {\fp F = \{[X1,1]\}, F = \{[X1,1],[X2,1]\}, ...}, whenever {\fp F} is unbound.
If subsequently a failure is detected, such as with the goal {\fp ran(F,\{1\}) \&
dom(F,\{\})}, then the computation loops forever and \setlog is not able to detect
the unsatisfiability.

Making the implementation of predicates over partial functions more
sophisticated as shown for instance in \citeN{CristiaRossiSEFM13} may help in
solving more efficiently a larger number of cases, but does not provide a
completely satisfactory solution in the general case. In fact, there are still
cases, such as that considered above, in which there is no simple finite
representation of the possibly infinite solutions and this may cause the
interpreter to go into infinite computations.

Most of the above mentioned problems could be solved by viewing partial
functions as first-class entities of the language and the operations dealing
with them as \emph{primitive constraints}, for which the constraint language
provides a suitable solver. Hence, the motivation for managing partial
functions through constraint solving is primarily to enhance the language
effectiveness, that is the ability to compute the
satisfiability/unsatisfiability of as many as possible (complex) set-based
formulas involving partial functions. Selecting \setlog as the host constraint
language for this embedding gives one the possibility to exploit its flexible
and general management of sets to represent partial functions and to provide
many basic set-theoretical operations on partial functions as primitive set
constraints for free. Other more specific operations on partial functions can be
added to the language as primitive constraints and the solver can be extended
accordingly.

The main original results of this work are: $(i)$ the identification of a small set
of operations on partial functions, to be dealt with as primitive constraints, which
are sufficient to represent all other common operations on partial functions as
simple conjunctions of these constraints; $(ii)$ the definition of a collection of
rewrite rules to simplify conjunctions of primitive constraints; $(iii)$ the
definition of a labeling mechanism based on the notion of finite representable
domains for partial functions; $(iv)$ the definition of a collection of inference
rules to detect possible inconsistencies without the need to perform time-consuming
labeling operations.

At our knowledge, only very few works have addressed the
problem of adding partial functions as primitive entities in a C(L)P setting. For
instance, the Conjunto language \cite{DBLP:journals/constraints/Gervet97} provides relation variables at the language
level. However, the domain and the range of the relations are limited to ground
finite sets. Map variables where the domain and range of the mapping can be also
finite set variables are introduced in CP(Map) \cite{Deville00}. All these proposals, however,
do not consider the more general case of partially specified partial
functions---where some elements of the domain or the range can be left
unknown---which on the contrary are essential in our proposal. Moreover, the
collection of primitive constraints on map variables they provide is usually
restricted to very few constraints, in particular to model the function application
operation.

The rest of this paper is organized as follows. In Section \ref{sec:clpset}, we
briefly recall the main features of the language \setlog. The new extended language
with partial functions is presented in Section \ref{sec:clppf}, focusing on what is
new with respect to \setlog. In Section \ref{sec:procedures} we describe the
constraint rewriting procedures for the new constraints and the global organization
of the constraint solver. The labeling mechanism with the introduction of pf-domains
is addressed in Section \ref{sec:pfdomains}. Section \ref{sec:improving} introduces a
number of inference rules that allow the solver to decide satisfiability of
irreducible constraints without having to resort to pf-domains, thus improving its
overall efficiency. A practical assessment of the performance of the new solver is
provided in Section \ref{empirical}.

\section{\{log\}}\label{sec:clpset}

\setlog is a \emph{Constraint Logic Programming (CLP)} language, whose constraint
domain is that of \emph{hereditarily finite sets}---i.e., finitely nested sets that
are finite at each level of nesting. \setlog allows sets to be \emph{nested} and
\emph{partially specified}---e.g., set elements can contain unbound variables, and it
is possible to operate with sets that have been only partially specified. \setlog
provides a collection of primitive constraint predicates, sufficient to represent all
the most commonly used set-theoretic operations---e.g., union, intersection,
difference.

The \setlog language was first presented by \citeN{Dovier04}. A complete constraint
solver for the pure CLP fragment included in \setlog---called \CLPSET---is described
by \citeN{Dovier00}, while its extension to incorporate intervals and Finite Domain
constraints is briefly presented by \citeN{Palu00}. Hereafter, with the name
\CLPSET\ we will refer to this last version of our constraint language, while \setlog
will refer to the whole language including \CLPSET, along with a number of other
syntactic extensions and extra-logical Prolog-like facilities. A working
implementation of \setlog (actually, an interpreter written in Prolog) is available
on the web \cite{SETLOG}.
%at \url{http://people.math.unipr.it/gianfranco.rossi/setlog.Home.html}.

\begin{comment}
\CLPSET is an instance of the general CLP scheme based on the
constraint domain ${\cal SET}$. The following are examples of terms denoting sets
(i.e., \emph{set terms}):
\begin{example}\label{ex:set terms}
Set terms:
\begin{itemize}
 \item $\{1,1,2\}$, $\{2,1\}$, and $\{1,2\}$,
 all denoting the same set composed of two elements, $1$ and $2$
 \item $\{X,\{\e,\{a,b\}\},\Int(1,10)\}$, denoting a set containing nested sets
 \item $\{X,Y|S\}$, denoting a partially specified set containing
 one or two elements, depending on whether $X$ is equal to $Y$ or not,
 and a, possibly empty, unknown part $S$.
\end{itemize}
\end{example}
\end{comment}

Sets are denoted by \emph{set terms}. For example, $\{1,1,2\}$, $\{2,1\}$, and
$\{1,2\}$ are set terms, all denoting the same set of two elements, $1$ and $2$;
$\{X,Y|S\}$ is a set term denoting a partially specified set containing one or two
elements, depending on whether $X$ is equal to $Y$ or not, and a, possibly empty,
unknown part $S$.

A \emph{primitive ${\cal SET}$-constraint} is defined as any literal based on the set
of predicate symbols $\Pi_C$ = $\{=,\In,\Cup,\disj,\leq,\Size,\set,\Integer\}$.
\begin{comment}
The predicates $=$ and $\In$ represent the equality and the membership relation,
respectively. The predicate $\Cup$ represents the union relation: $\Cup(r,s,t)$ holds
if and only if $t = r \cup s$. The predicate $\disj$ represents the disjoint
relationship between two sets: $\disj(s,t)$ holds if and only if $s \cap t = \e$. The
predicate $\Size$ represents set cardinality: $\Size(s,n)$ holds if and only if $n =
|s|$. Finally, the predicate $\leq$ represents the comparison relation ``less or
equal'' over the integer numbers. Predicate symbols $\Neq$, $\Nin$, $\nun$,
$\ninteger$, \dots, are used to denote the negated versions of the corresponding
constraint predicates---e.g., $s \Nin t$ represents the literal $\neg \: ( s \In t
)$.
\end{comment}
Symbols in $\Pi_C$ have their natural set-theoretic interpretation. In particular,
the predicate $\Cup$ represents the union relation ($\Cup(r,s,t)$ holds if and only
if $t = r \cup s$), while the predicate $\disj$ represents the disjoint relation
between two sets ($\disj(s,t)$ holds if and only if $s \cap t = \e$). Most other
useful set-theoretical predicates, e.g., $\Subseteq$ and $\Cap$, can be defined as
${\cal SET}$-constraints, using $\disj$ and $\Cup$---e.g., $\Subseteq(u,v)
\Leftrightarrow \Cup(u,v,v)$ \cite{Dovier00}. As an example, the following formula,
$\Cap(R,S,T) \wedge \Size(T,N) \wedge N =< 2$, is an admissible ${\cal
SET}$-constraint whose (informal) interpretation is: the cardinality of $R \cap S$
must be not greater than $2$.

\begin{comment}
\begin{example}\label{ex:setlog-constraint}
The following formulas are admissible ${\cal SET}$-constraints ($R$, $S$, $T$, $X$,
and $N$ are variables):
\begin{itemize}
 \item[$(i)$] $1 \In R \wedge 1 \Nin S \wedge \Cap(R,S,T) \wedge T = \{X\}$
 \item[$(ii)$] $\Cap(R,S,T) \wedge \Size(T,N) \wedge N =< 2$.
\end{itemize}
Their informal interpretation is as follows: $(i)$ the set $T$ is the intersection
between sets $R$ and $S$, $R$ must contain $1$ and $S$ must not, and $T$ must be a
singleton set; $(ii)$ the cardinality of $R \cap S$ must be not greater than $2$.
\end{example}

Similarly, other interesting integer predicates (e.g., $<$, $\geq$, and $>$) can be
defined as ${\cal SET}$-constraints using $\leq$ and $=$.
\end{comment}

\CLPSET is endowed with a complete \emph{constraint solver}, called $\SAT$, for
verifying the satisfiability of ${\cal SET}$-constraints. Given a constraint $C$,
$\SAT(C)$ transforms $C$ either to \false (if $C$ is unsatisfiable) or to a finite
collection $\{C_1,$ $\dots,$ $C_k\}$ of constraints in {\em solved form}. A constraint in solved form is guaranteed to be satisfiable w.r.t.
the underlying interpretation structure.
%${\cal A}_{\cal SET}$
Moreover, the disjunction of all the constraints in solved form generated by
$\SAT(C)$ is equisatisfiable to $C$ in %${\cal A}_{\cal SET}$.
the structure. A detailed description of the constraint solver $\SAT$ can be found in
\citeN{Dovier00}.

\begin{example}\label{ex:solving solved form}
Let $C$ be $\{1,2\,|\,X\} = \{1\,|\,Y\} \wedge 2 \Nin X$. Then $\SAT(C)$ returns, one
by one, the following three answers, each of which is a constraint in solved form: $Y
= \{2\,|\,X\} \wedge 2 \Nin X \wedge \set(X)$; $X = \{1\,|\,N\} \wedge Y =
\{2\,|\,N\} \wedge \set(N) \wedge 2 \Nin N$; and $Y = \{1,2\,|\,X\} \wedge 2 \Nin X
\wedge \set(X)$ (where $N$ is a new variable).
%Observe that: ${\cal A}_{\cal SET} \models \forall X \forall Y\, (C \leftrightarrow
%\exists N\,(C_1 \vee C_2 \vee C_3))$.
\end{example}

\section{The extended language \CLPPF}\label{sec:clppf}

The constraint domain ${\cal SET}$ is extended so as to incorporate partial
functions. The new constraint domain and the related language are called ${\cal PF}$
and \CLPPF, respectively. Since ${\cal PF}$ includes ${\cal SET}$ as a special case
we will simply highlight what is new in ${\cal PF}$ with respect to ${\cal SET}$.

As concerns syntax, our choice is to not introduce any special symbol to
represent partial functions, since they can be easily represented as sets.
Partial functions are just a particular kind of sets. Forcing a set to represent
a partial function will be obtained at run-time by using suitable constraints on
its elements.

\begin{definition}
We say that a set term $r$ represents a \emph{partial function} if $r$ has one
of the forms: $\{\}$ or $\{[x_1,t_1],[x_2,t_2],\dots,[x_n,t_n]\}$ or
$\{[x_1,t_1],[x_2,t_2],\dots,[x_n,t_n] \mid s\}$, and $x_i$, $t_i$, $i =
1,\dots,n$, are terms, $s$ is a set term representing a partial
function, and the constraints $x_i \neq x_j$, $x_i \not\in \dom s$, hold for
all $i, j = 1,\dots,n$, $i \neq j$.
\end{definition}

A critical issue in the definition of ${\cal PF}$ is the
choice of which operations over partial functions should be {\em
primitive\/}---i.e., part of $\Pi_C$\/---and which, on the contrary, should be
{\em programmed\/} using the language itself. Minimizing the number of predicate
symbols in $\Pi_C$ has the advantage of reducing the number of different kinds
of constraints to be dealt with and, hopefully, simplifying the language and its
implementation. On the other hand, having to implement such operations on top of
the language may lead to efficiency and effectiveness problems, similar to those
encountered with the implementation of partial functions using \setlog discussed
in Section~\ref{sec:introduction}.

Our choice is to extend the set $\Pi_C$ of constraint predicate symbols with the
following four predicate symbols:
\[
\Dom, \Ran, \Comp, \Pfun
\]
The intuitive interpretation of these predicate symbols is: $\Dom(r,a)$ (resp.
$\Ran(r,a)$) holds iff $a$ is the domain (resp., range) of the partial function
$r$; $\Comp(r,s,t)$ holds iff the partial function $t$ is the composition of the
partial functions $r$ and $s$, i.e. $t = \{[x,z] : \exists y([x,y] \in r \land
[y,z] \in s)\}$; $\Pfun(r)$ holds iff $r$ is a partial function.

Atomic predicates based on these symbols are the only primitive constraints that
\CLPPF offers to deal with partial functions (let us simply call these constraints
\emph{${\cal PF}$-constraints}). A (general) \emph{$({\cal SET,PF})$-constraint} is
just a conjunction of primitive constraints built using the enlarged $\Pi_C$, i.e.
$\{=,\In,\Cup,\disj,\leq,\Size,\set,\Integer\} \cup \{\Dom, \Ran, \Comp, \Pfun\}$.

The following theorem ensures that the primitive constraints are
sufficient to define most of the common operations on partial functions as $({\cal
SET,PF})$-constraints. Complete proofs of this and the remaining theorems are
available on-line at
\url{http://people.math.unipr.it/gianfranco.rossi/SETLOG/setlogpf_proofs.pdf}.
 Many of these theorems were formally proved using the Z/EVES proof assistant
\cite{ZMT}.

\begin{theorem}\label{equivalent conjs}
Literals based on predicate symbols: $\Dres$ (domain restriction), $\Rres$ (range
restriction), $\Ndres$ (domain anti-restriction), $\Nrres$ (range anti-restriction),
$\Rimg$ (relational image), $\Oplus$ (overriding) and $\Id$ (identity) can be
replaced by equivalent conjunctions of literals based on $=$, $\Cup$, $\disj$,
$\Dom$, $\Ran$ and $\Comp$.
\end{theorem}

\begin{proof*}[Proof (sketch)]
The following equivalences hold:
%\emph{Proof [sketch]}
$$\begin{array}{rcl}
\Ndres(a,r,s) & \iff  & \Dres(a,r,b) \land \cdiff(r,b,s) \\
\Nrres(b,r,s) & \iff & \Rres(b,r,a) \land \cdiff(r,a,s)  \\
\Dres(a,r,s) & \iff & \Dom(r,dr) \land \Dom(s,ds) \land \Cap(a,dr,ds) \land
\Subseteq(s,r) \\
\Rres(b,r,s) & \iff & \Cup(s,t,r) \land \Ran(s,rs) \land \Ran(r,rr) \\
             &       & \land\ \Cap(b,rr,rs) \land \Ran(t,rt) \land \disj(rs,rt) \\
\Rimg(b,r,s) & \iff & \Dres(b,r,rb) \land \Ran(rb,s) \\
\Oplus(r,s,t) & \iff & \Cup(rs,s,t) \land \Ndres(ds,r,rs) \land \Dom(s,ds) \\
\Id(a,r) & \iff & \Dom(r,a) \land \Ran(r,a) \land \Comp(r,r,r) \mathproofbox
\end{array}
$$
\end{proof*}

Other common operations on partial functions can be defined in the same way. For
example, the application of a partial function $r$ to an element $x$ can be easily
defined in terms of primitive constraints as follows: $\Apply(r,x,y)$ is true if and
only if $[x,y] \In r$ holds.

The ability to express operations on partial functions as $({\cal
SET,PF})$-constraints as stated in Theorem \ref{equivalent conjs} allows us to not
consider these operations in the definition of the constraint solver for $\CLPPF$ and
to focus our attention only on the four primitive constraints based on $\Pfun$,
$\Dom$, $\Ran$ and $\Comp$.

It is worth noting that the proposed subset of primitive predicate symbols
is not the only possible choice. Roughly speaking, it is motivated by observing
that: since a function is a tuple of the form $(dom, law, ran)$, then choosing
$\Dom$ and $\Ran$ seems a rather natural choice; the $law$ can be given as
membership predicates (i.e. $\apply$) which is already part of the primitive
constraints; $\Pfun$ is easy to justify since it is necessary to state which sets
are partial functions; finally, $\Comp$ is justified by observing that it is
hardly definable in terms of the other primitive constraints. However, proving
that this subset of primitive constraints is the minimal one, as well as
comparing our choice with other possible choices, in terms of, e.g., expressive
power, completeness, effectiveness, and efficiency, is out of the scope of the
present work.

\section{Constraint Rewriting Procedures} \label{sec:procedures}

For each primitive constraint symbol $\pi \in \Pi_C$, we develop a \emph{constraint
rewriting procedure} specifically devoted to process that type of constraint.
Basically, each procedure repeatedly applies to the input constraint $C$ a collection
of \emph{rewrite rules} for $\pi$ until either $C$ becomes \false or no rule for
$\pi$ applies to $C$. At any moment, $C$ represents the \emph{constraint store}
managed by the solver.

The rewrite rules have the following general form
$$
\frac{\textit{pre-conditions}}{\{C_1,\dots,C_n\} \rightarrow
        \{C_1',\dots,C_m'\}}
$$
\noindent where $C_i$ and $C_i'$ are primitive $({\cal SET,PF})$-constraints and
\textit{pre-conditions} are (possibly empty) boolean conditions on the terms
occurring in $C_1,\dots,C_n$. In order to apply the rule, all \textit{pre-conditions}
need to be satisfied. $\{C_1,\dots,C_n\} \rightarrow \{C_1',\dots,C_m'\}$ ($n$, $m$
$\geq 0$) represents the changes in the constraint store caused by the rule
application.

Some rewrite rules for dealing with single ${\cal PF}$-constraints are shown in
Figures \ref{f:domrules} and \ref{f:comprules}; all of them can be found in the
online appendix (Appendix A).
%at \url{http://people.math.unipr.it/gianfranco.rossi/SETLOG/setlogpf_rules.pdf}.
%Figures \ref{fig:rules_dom}---\ref{fig:rules_pfun}.
Rewrite rules for all other primitive constraints can be found elsewhere \cite{Dovier00,Palu00}.

\begin{figure}
\figrule
\begin{gather*}
\frac{r \in \mathcal{V}}{\{\Dom(r,r)\} \rightarrow \{r = \e\} } \\[1mm] \label{rule:dom_emptyset2}
\frac{\E(a)}
 {\{\Dom(r,a)\} \rightarrow \{r = \e\} } \\[1mm] \label{rule:dom_emptyset1}
\frac{\E(r)}
 {\{\Dom(r,a)\} \rightarrow \{a = \e\} } \\[1mm] \label{rule:dom_knownset}
\frac{r = \{[x,y]|rr\} \phantom{aaa} \lnot \E(a)}
 {\{\Dom(r,a)\} \rightarrow {} \{a = \{x | rs\}, [x,y] \Nin rr, \Dom(rr,rs)\} }
 \\[1mm]
\label{rule:dom_var}
\frac{r \in \mathcal{V} \phantom{aaa} a = \{x | rs\} }
 {\{\Dom(r,a)\} \rightarrow \{r = \{[x,y]|rr\}, x \Nin rs, \Dom(rr,rs)\} }
\end{gather*}
\caption{\label{f:domrules}Rewrite rules for $\Dom$.}
\figrule
\begin{gather*}
\frac{\E(q) \phantom{aaa} \lnot\E(r) \phantom{aaa} \lnot\E(s)}
  {\{\Comp(r,s,q)\} \rightarrow \{\Ran(r,rr), \Dom(s,ds), \disj(rr,ds)\}}
  \label{rule:comp_emptyset} \\[1mm]
\frac{q = \{[x,z]|rq\} \phantom{aaa} \lnot\E(r) \phantom{aaa} \lnot\E(s)}
  {\begin{array}{c}
    \{\Comp(r,s,q)\} \rightarrow \{r = \{[x,y] | rr\},  \\
    s = \{[y,z] | rs\}, [x,z] \Nin rq, [y,z] \Nin rs, \Comp(rr,s,rq)\}
  \end{array}
  } \label{rule:comp_qset} \\[1mm]
\frac{q \in \mathcal{V} \phantom{aaa} r = \{[x,y] | rr\} \phantom{aaa}
\lnot\E(s) \phantom{aaa} s \notin \mathcal{V}}
  {\begin{array}{c}
    \{\Comp(r,s,q)\} \rightarrow \{s = \{[y,z] | rs\}, \\
     q = \{[x,z] | rq\}, [x,y] \Nin rr, [y,z] \Nin rs, \Comp(rr,s,rq)\} \\
    \mathsf{or} \\
    \{\Comp(r,s,q)\} \rightarrow \{\Dom(s,ds),  y \Nin ds,  [x,y] \Nin rr, \\
     \Comp(rr,s,q)\}
   \end{array}
 } \label{rule:comp_qvar}
\end{gather*}
\caption{\label{f:comprules}Rewrite rules for $\Comp$.}
\figrule
\end{figure}

\begin{comment}
Rewrite rules \ref{rule:comp_qvar} and \ref{rule:pfun_nonground} are nondeterministic
rules: if the preconditions are met, the rule nondeterministically performs one of
the two rewritings in its lower part. In particular, rule \ref{rule:comp_qvar} deals
with the case in which both $r$ and $s$ in $\Comp(r,s,q)$ are not variables nor empty
partial functions. The nondeterministic choice takes care of the fact that, for each
pair $[x,y]$ in $r$, it may exist a $z$ such that $[y,z] \in s$ or it may not exist
any $z$ such that $[y,z] \in s$. This last condition is expressed by stating that
$\Dom(s,ds) \wedge  y \Nin ds$. Instead, in rule \ref{rule:pfun_nonground} the
nondeterministic choice accounts for the fact that a pair $[x,y]$ can occur exactly
once in $r$ or it may occur more than once in $r$ and in the last case
these multiple occurrences of $[x,y]$ must be simply ignored. %since $r$ is a set.

Observe that not all rules are strictly necessary. For instance, rule
\ref{rule:pfun_ground} could always be replaced by the more general rule
\ref{rule:pfun_nonground}. However, we prefer to single out special cases (e.g. the
one dealt by rule \ref{rule:pfun_ground}) in order to manage them in a more efficient
way.
\end{comment}

The global organization of the solver for the new language---called $\SATPF$---is
shown in Algorithm \ref{glob}. It makes use of two procedures: {\sf infer} and {\sf
STEP}. \textsf{infer} is used to automatically add the constraints $\set$,
$\Integer$, and $\Pfun$ to the constraint $C$ to force arguments of primitive
constraints to be of the proper type. For example, if $C$ contains the constraint
$\Dom(r,a)$ then ${\sf infer}(C)$ will add to $C$ the constraint $\Pfun(r) \wedge
\set(a)$.
%Executing procedure {\sf infer} before entering the rewriting loop (procedure {\sf
%STEP}) ensures that if, for instance, $\Dom(r,a)$ is in the constraint store then
%also $\Pfun(r)$ and $\set(a)$ are in the constraint store.
The procedure \textsf{STEP} is the core part of $\SATPF$: it applies specialized
constraint rewriting procedures to the current constraint $C$ and returns the
modified constraint. The execution of $\textsf{STEP}$ is iterated until a fixpoint is
reached---i.e., the constraint cannot be simplified any further. Notice that {\sf STEP} returns $\false$ whenever
(at least) one of the procedures in it rewrites $C$ to $\false$. Moreover, {\sf
STEP($\false$)} returns $\false$.

\begin{algorithm}[htbp]
\begin{algorithmic}[0]
 \Procedure{$\SATPF$}{$C$}
 \State $C \gets \textsf{infer}(C)$
 \Repeat
 \State $C' \gets C$;
 \State $C \gets \textsf{STEP}(C)$;
 \Until{$C = C'$;}
 \State\Return{$C$}
 \EndProcedure
\end{algorithmic}
\caption{The \CLPPF Constraint Solver} \label{glob}
\end{algorithm}

When no rewrite rule applies to the considered ${\cal PF}$-constraint
%(e.g. whenever $r$ and $s$ in $\Dom(r,s)$ are both variables and they are not the
%same variable)
then the corresponding rewriting procedure terminates immediately and the constraint
store remains unchanged. Since no other rewriting procedure deals with the same kind
of constraints, the irreducible constraints will be returned as part of the
constraint computed by $\SATPF$. Precisely, if $X$ and $X_i$ are variables and $t$ is
a term (either a variable or not), the following ${\cal PF}$-constraints are dealt with
as \emph{irreducible}:
\begin{enumerate}
\item $\Dom(X_1,X_2)$, where $X_1$ and $X_2$ are distinct variables;
\item $\Ran(X,t)$, where $t$ is distinct from $X$ and $t$ is not the empty set;
\item $\Comp(X_1,t,X_3)$ or $\Comp(t,X_2,X_3)$, where $t$ is not the empty set;
\item $\Pfun(X)$ and there are no constraints of the form
$\Integer(X)$ in $C$.
\end{enumerate}

Roughly speaking, the irreducible constraints are these because we are
not able to rewrite them to \emph{finite} conjunctions of primitive $({\cal
SET,PF})$-constraints. In particular, solving the constraint $\Ran(X,t)$, where
$t$ is a set term not denoting the empty set, would amount to solve the formula
$\forall x(x \in X \iff \exists y,z (x = [y,z] \wedge z \in t))$ which is not
expressible as a finite conjunction of primitive $({\cal SET,PF})$-constraints.
Notice that, conversely, the case $\Dom(X,t)$, where $t$ is a set term (e.g.
$\Dom(X,\{1\})$), can be easily rewritten to a finite conjunction of primitive
constraints since the cardinality of $X$ is necessarily that of $t$; hence this
constraint is not dealt with as irreducible.

For all other primitive $({\cal SET,PF})$-constraints, $\SATPF$ uses the rewriting
rules of \CLPSET\ and the irreducible form constraints it returns are all ${\cal
SET}$-constraints in solved form (cf. Sect. \ref{sec:clpset} and \citeN{Dovier00}).
Observe that, a constraint composed of only solved form literals is proved to be
always satisfiable.
%with respect to the underlying interpretation structure ${\cal A}_{\cal SET}$, hence
%with respect to ${\cal A}_{\cal PF}$.

%A constraint $C$ is in irreducible form if it is empty or if all its components are
%simultaneously in irreducible form.

\begin{example} Constraint rewriting.
\begin{itemize}
\item $\Dom(\{[a,1],[b,2],[c,1]\},D)$ is rewritten to $D = \{a,b,c\}$

\item $\Dom(\{[a,1]\},\{b\})$ is rewritten to \false

\item $\Comp(\{[1,b]\},B,\{[1,a]\})$ is rewritten to $B = \{[b, a] | BR\} \land
[b,a] \Nin BR \land \Pfun(BR) \land \Dom(BR,D) \land b \Nin D \land \set(D)$

\item $\Cap(\{X\},\{1\},D) \land \Dom(R,D) \land \Ran(R,\e)$ is rewritten to
$D = \e \land R = \e \land X \Neq 1$

\item $\Apply(F,X,Y) \land \Dom(F,D) \land X \Nin D$ is rewritten to \false.
\end{itemize}
\end{example}

Note that with the implementation of $\Dom$ and $\Ran$ as user-defined \setlog{}
predicates (see \citeN{CristiaRossiSEFM13}) the last goal would loop forever.

The $\SATPF$ procedure is proved to be always terminating.

\begin{theorem}[Termination]\label{termination-glob}
The $\SATPF$ procedure terminates for every input constraint $C$.
\end{theorem}

The termination of $\SATPF$ and the finiteness of the number of non-deterministic
choices generated during its computation guarantee the finiteness of the number of
constraints non-deterministically returned by $\SATPF$. Therefore, $\SATPF$\ applied
to a constraint $C$ always terminates, returning either $\false$ or a (finite)
disjunction of $({\cal SET,PF})$-constraints in a simplified form. The following
theorem proves that the collection of constraints in irreducible form generated by
$\SATPF$ preserves the set of solutions of the input constraint, hence, it is
correct.
%this disjunction is proved to be logically equivalent to the given constraint $C$

\begin{theorem}[Equisatisfiability]\label{sound}
Let $C$ be a constraint, $C_{1},$ $\ldots,$ $C_n$ be the constraints obtained from
$\SATPF(C)$, $\sigma$ be a valuation of $C$ and $C_{1} \vee \ldots \vee C_n$,
expanded to the new variables possibly introduced into $C_{1},\ldots,C_n$ by the
rewrite procedures, and ${\cal A}_{\cal PF}$ be the interpretation structure
associated with the constraint domain ${\cal PF}$. Then, ${\cal A}_{\cal PF} \models
\sigma(C)$ if and only if ${\cal A}_{\cal PF} \models \sigma(C_{1} \vee \ldots \vee
C_n)$.
\end{theorem}

If at least one of the constraints $C_{i}$ returned by $\SATPF(C)$ contains
\emph{only} primitive $\cal SET$-constraints then, according to \citeN{Dovier00},
$C_{i}$ is in solved form and it is surely satisfiable. Therefore, in this case,
thanks to Theorems \ref{termination-glob} and \ref{sound}, we can conclude that the
original constraint $C$ is surely satisfiable.

Unfortunately, this is not always the case, as discussed in the next section.

\section{pf-domains}\label{sec:pfdomains}

Differently from  \CLPSET, the simplified constraint returned by $\SATPF$ is not
guaranteed to be satisfiable.

\begin{example}\label{ex:unsatisfiable}
The following $({\cal SET,PF})$-constraint
\[
\Dom(R,D) \wedge R \Neq \e \wedge \Cup(D,Y,Z) \wedge \disj(D,Z)
\]
is an irreducible constraint but it is clearly unsatisfiable (the only
possible solution for $\Cup(D,Y,Z) \wedge \disj(D,Z)$ is $D = \e$, but $D = \e$ if
and only if $R = \e$).
\end{example}

Thus, differently from \CLPSET, the ability to produce a collection of constraints in
an irreducible form from the input constraint $C$ cannot be used to decide the
satisfiability of $C$. As many concrete solvers, e.g. the \CLPFD solvers, $\SATPF$ is
an \emph{incomplete} solver. Thus, if it returns \false\ the input constraint is
surely unsatisfiable, whereas if it returns a constraint in irreducible form then we
cannot conclude that the input constraint is surely satisfiable.

In order to obtain a complete solver, we provide a way to associate a
\emph{finitely representable domain} to each partial function variable and to
force these variables to get values from their associated domains, i.e. to
perform \emph{labeling} on them. This is obtained by defining a new primitive
constraint $\Pfun$, of arity $2$, with the following interpretation: $\Pfun(r,n)
\text{ holds if and only if }
 r \in X \pfun Y \wedge n \in \nat \wedge \vert r \vert \leq n$.

The solutions of $\Pfun(r,n)$ are all the partial functions $r$ with cardinality
less than or equal to $n$. The ability to represent domains and ranges of partial
functions as partially specified sets, i.e. sets containing unbound variables as
their elements, allows us to provide a \emph{finite} representation for the
(possibly infinite) set of all solutions of $\Pfun(r,n)$. For example, the set
of solutions for $\Pfun(r,2)$, where $r$ is a variable, can be represented by
the following equisatisfiable disjunction of three primitive constraints: $r =
\e \lor r = \{[X,Y]\} \lor r = \{[X_1,Y_1],[X_2,Y_2]\} \wedge X_1 \Neq X_2$.

We will call the set of partial functions represented by these constraints the
\emph{pf-domain} of the pf-variable $r$. pf-domains represent in general
infinite sets but they are finitely representable in our language.

>From an operational point of view, solving $\Pfun(r,n)$, with $n$ a constant
natural number, non-deterministically computes, one after the other, all the $n
+ 1$ possible assignments for $r$. Therefore, solving  $\Pfun(r,n)$ allows us to
perform a sort of \emph{labeling} over the pf-variable $r$. Note that, differently from $\Pfun(r)$, $\Pfun(r,n)$ has no
irreducible form. If $r$ is an unbound variable ($n$ is required to be a constant
number), then solving $\Pfun(r,n)$ always generates an equality for $r$, along with
possible inequality constraints over the elements in the domain of $r$.

The labeling process involved in $\Pfun/2$ constraints do not compromise termination
of the procedure $\SATPF$ since the set of possible values to be assigned to partial
function variables through labeling is anyway finite. Moreover, assuming our domain of
discourse is limited to finite partial functions, it is straightforward to see
that the rewriting rules for $\Pfun/2$ preserve the set of solutions of the input
constraint. Thus we can immediately extend to $\Pfun/2$ constraints the results of
Theorems \ref{termination-glob} and \ref{sound}.

Solving $\Pfun/2$ constraints allows pf-variables to always get a value, although it
can be a non-ground value. This is enough, however, to guarantee that all ${\cal
PF}$-constraints are completely eliminated at the end of the computation.

\begin{lemma}\label{elimination}
Let $C$ be an input constraint and $V_1,\dots,V_n$ all the pf-variables occurring in
$C$. If $C$ contains $\Pfun(V_1,k_1) \wedge \dots \wedge \Pfun(V_n,k_n)$,
$k_1,\dots,k_n \in \nat$, then $\SATPF(C)$ returns either \false or a disjunction
of ${\cal SET}$-constraints in \emph{solved form}.
\end{lemma}

Remembering that $\cal SET$-constraints in solved form are always satisfiable, Lemma
\ref{elimination} guarantees that, if the input constraint $C$ contains $\Pfun/2$
constraints for all the pf-variables occurring in it and $\SATPF(C)$ does not
terminate with \false, then the disjunction of constraints returned by $\SATPF(C)$ is
surely satisfiable. Since $\SATPF$ is proved to preserve the set of solutions of
$C$ (cf. Theorem \ref{sound}), then we can conclude that in this case $C$ is
satisfiable.

Hence, by properly exploiting $\Pfun/2$ constraints, we get a complete solver. This
means that, once $k_1,\dots,k_n$ are fixed, our solver can detect all cases in
which the input constraint is unsatisfiable, as well as all cases in which the input
constraint is satisfiable and, in these cases, it can generate all viable solutions.

\begin{example} %Using pf-domains.
The following constraints are rewritten to either \false or to a solved form
constraint, whereas they are left unchanged if no pf-domain is specified.
\begin{itemize}
\item $\Dom(R,D) \wedge D \Neq \e \wedge \Cup(D,Y,Z) \wedge \disj(D,Z) \wedge
\Pfun(R,5)$ is rewritten to \false

\item $\Ran(X,\{1\}) \wedge \Cup(X,Y,Z) \wedge \Pfun(X,5)$ is rewritten to
the solved form constraint (first solution): $X = \{[A,1]\} \wedge Z = \{[A,1] | Y\}
\wedge \set(Y)$.

\begin{comment}
\item  $\Ran(Z1,R) \wedge \Ran(Z2,R) \wedge \Dom(Z1,S) \wedge \Dom(Z2,S) \wedge
Z1 \Neq Z2 \wedge \Pfun(Z1,100)$ is rewritten to the solved form constraint (first
solution): $Z1 = \{[A,B],[C,D]\} \wedge Z2 = \{[A,D],[C,B]\} \wedge R = \{B,D\}
\wedge S = \{A,C\} \wedge B \Neq D \wedge A \Neq C$

\item $\Comp(\{[1,a]\},Y,Z) \wedge \Dom(Y,S) \wedge a \Nin S \wedge Z \Neq \e \wedge
\Pfun(Y,5)$ is rewritten to \false.
\end{comment}

\end{itemize}
\end{example}

\begin{comment}
Differently from proposals dealing with Finite Domains our constraint solver does not
perform any domain reduction nor any propagation due to domain changes to all the
involved constraint variables. Conversely, in our solver pf-domains are exploited
uniquely to perform labeling over pf-variables, that is to bound pf-variables to
terms that represent possible values for them.
\end{comment}

\begin{comment}
Finally, it is worth noting that, while having to specify labeling information
through the $\Pfun/2$ constraints may be cumbersome in some cases, it can be of great
importance in other cases. For example, in the application described in
\cite{CristiaSEFM13} where \setlog{} is used as a test case generator it may be
important to be able to generate possible values (actually, just one solution is
enough in this case) for the variables occurring in goals that are proved to be
satisfiable.
\end{comment}

\section{Improving constraint solving}\label{sec:improving}

>From a more practical point of view, having to perform labeling for pf-variables
may cause unacceptable execution time in some cases. For example, the constraint
\[
\Dom(R,D1) \wedge \Dom(R,D2) \wedge D1 \Neq D2 \wedge \Pfun(R,k)
\]
is proved to be unsatisfiable, but only for relatively small values of $k$.

To alleviate this problem, we introduce a number of new rewrite rules---hereafter
simply called \emph{inference rules}---that allow new constraints to be inferred from
the irreducible constraints.
%left by the solver in the constraint store.
The presence of these additional constraints allows the solver to deduce possible
unsatisfiability of the given constraint without having to resort to any labeling
process, thus improving the overall efficiency of constraint solving in many cases.

The inference rules are applied by calling function \textsf{infer\_rules} just after
the iteration of $\textsf{STEP}$ ends finding a fixpoint (see Algorithm \ref{glob}).
\textsf{infer\_rules}($C$) applies all possible inference rules to all possible
primitive constraints in $C$. After the rules have been applied, possibly modifying
$C$, the $\textsf{STEP}$ loop is repeated from the beginning. Only when both
$\textsf{STEP}$ and \textsf{infer\_rules} do not modify $C$, then the new global
constraint solving procedure---called $\SATPF'$---ends.

Some of the inference rules used by $\SATPF'$ are shown in Figure
\ref{f:infrules}; all of them can be found in the online appendix (Appendix A).
%at \url{http://people.math.unipr.it/gianfranco.rossi/SETLOG/setlogpf_rules.pdf}.
Each inference rule captures some property of the
primitive operators for partial functions, possibly relating these operators with
other general operators, such as inequality (constraint $\Neq$) and set cardinality
(constraint $\Size$). All rules take into account one or two primitive constraints at
a time and add new primitive constraints to the constraint store.

\begin{figure}
\figrule
\begin{gather}
\frac{}
 {\{\Dom(r,a), \Dom(r,b)\} \rightarrow \{\Dom(r,a), a = b\} }
 \label{irule:dom_dom} \\[1mm]
\frac{a \in \mathcal{V}}
 {\{\Ran(r,a), r \Neq \e\} \rightarrow \{\Ran(r,a), r \Neq \e, a \Neq \e\} }
 \label{irule:ran_neq2} \\[1mm]
%\frac{r \in \mathcal{V} & a \in \mathcal{V} & n \in \mathcal{V}}
\frac{}
 {\{\Dom(r,a)\} \rightarrow \{\Dom(r,a), \Size(r,n), \Size(a,n)\} }
 \label{irule:dom_size1} \\[1mm]
\frac{}
 {\begin{array}{cl}
 \{\Comp(r,s,q)\} \rightarrow & \{\Comp(r,s,q), \\
                              & \phantom{a} \Dom(q,a), \Dom(r,b), \Subseteq(a,b)\}
 \end{array}
 }
 \label{irule:comp_dom_ran1} \\[1mm]
\frac{}
 {\begin{array}{cl}
%  \{\Cup(r,s,q), \Pfun(r), \Pfun(s), \Pfun(q)\} \rightarrow \\
  \{\Cup(r,s,q), \Pfun(q)\} \rightarrow & \{\Cup(r,s,q), \Pfun(q),  \\
       & \phantom{a} \Dom(r,dr), \Dom(s,ds), \Dom(q,dq), \\
       & \phantom{a} \Cup(dr,ds,dq) \}
 \end{array} }
 \label{irule:un_dom}
\end{gather}
\caption{\label{f:infrules}Some inference rules.}
\figrule
\end{figure}
\begin{example} \label{ex:unsatisfiable2}
The following constraints are all proved to be unsatisfiable using $\SATPF'$
(see the applied rules in Figure \ref{f:infrules}):
%, whereas their unsatisfiability is not detected using $\SATPF$, that is without
%applying any inference rule (in this case they are simply treated as irreducible).
\[
\begin{array}{ll}

\Dom(X,D1) \wedge \Dom(X,D2) \wedge D1 \Neq D2 & \text{(rule
\eqref{irule:dom_dom})} \\

\Ran(X,RX) \wedge RX \Neq \e \wedge \disj(X,Z) \wedge \Cup(X,Y,Z) & \text{(rule
\eqref{irule:ran_neq2})}\\

\Dom(X,DX) \wedge \Size(X,N) \wedge \Size(DX,M) \wedge N \Neq M & \text{(rule
\eqref{irule:dom_size1})} \\

\Comp(\{[a,1]\},Y,Z) \wedge \Dom(Z,DZ) \wedge a \Nin DZ \wedge Z \Neq \e
 & \text{(rule \eqref{irule:comp_dom_ran1})} \\

\Cup(X,Y,Z) \wedge \Dom(X,D) \wedge \Dom(Y,D) \wedge \Dom(Z,DZ) \wedge D \Neq DZ
 & \text{(rule \eqref{irule:un_dom})} \\

\end{array}
\]
\end{example}

The same constraints of Example \ref{ex:unsatisfiable2} but using $\SATPF$, that is
without applying any inference rule, are simply treated as irreducible. On the other
hand, adding constraints $\Pfun$/2 to perform labeling on pf-variables would allow
$\SATPF$ to detect the unsatisfiability for all these constraints, but only when the
specified partial function cardinalities are relatively small the response times
would be practically acceptable.

Termination of the improved constraint solver is stated by the following theorem.

\begin{theorem}[Termination of $\SATPF'$]\label{termination-glob2} The
$\SATPF'$ procedure can be implemented in such a way that it terminates for every
input constraint $C$.
\end{theorem}

Soundness of the extended solver $\SATPF'$ comes from soundness of $\SATPF$ and from the following theorem, which ensures that the added
constraints do not modify the set of solutions of the original constraint.

\begin{theorem}[Equisatisfiability of inference rules]\label{equisatisfiable}
Let $S$ be a constraint and $S'$ be the constraint obtained from the inference rules. Then $S'$ is equisatisfiable to $S$ with
respect to the interpretation structure ${\cal A}_{\cal PF}$.
\end{theorem}

%It is worth noting that, at the implementation level, the solver guarantees that
%inference rules are applied only when it is really useful. For example, rule
%\ref{irule:dom_size1} is applied only if the constraint store already contains some
%constraint of the form  $\Size(r,k)$ or $\Size(s,k)$, which are necessary for the
%solver to find out possible inconsistencies. Thus, even that no preconditions are
%explicitly stated for the rule, in practice applying the rule requires to consider
%more than one constraint at a time in the constraint store. Moreover, note that
%labeling, on both integer and pf-variables, is postponed as much as possible. This
%allows the solver first to prune the search space using the more efficient
%deterministic rewrite rules and only at the end of the computation to generate
%possible values for the unbound variables.

$\SATPF'$ is still not a complete solver unless $\Pfun/2$ is used for all
pf-variables. As a counterexample, consider the following constraint
\[
\Ran(X,\{1\}) \wedge \Ran(Y,\{1,2\}) \wedge \Dom(X,D) \wedge \Dom(Y,D) \wedge
\disj(X,Y).
\]
This constraint is unsatisfiable with respect to ${\cal A}_{\cal PF}$, but $\SATPF'$
is not able to prove this fact (it simply leaves the constraint unchanged).

New inference rules could be added to the solver to detect further
properties of the partial function domain, thus avoiding as much as possible the need
for $\Pfun$/2 constraints.
\begin{comment}
For example, the following inference rule relates $\Comp$, $\Size$ and
$\Ran$:
%\frac{s \in \mathcal{V} & q \in \mathcal{V} & \lnot\E(r) & \lnot r \in \mathcal{V}}
$$\frac{}
 {\{\Comp(r,s,q)\} \rightarrow \{\Comp(r,s,q), \Ran(q,a), \Size(a,n), \Size(s,m), n \leq m\}
 }$$
\end{comment}
However, finding a collection of inference rules that guarantees to obtain a
complete solver, regardless of the presence of $\Pfun$/2 constraints, seems to be a
difficult task. Moreover, checking the constraint store to detect applicable
inference rules may be quite costly in general. Thus, the solution we adopted is
based on finding a tradeoff between efficiency and completeness, as usual in many
concrete constraint solvers. Only those properties that require relatively small
effort to be checked are taken into account by the solver. For all cases not covered
by the inference rules, however, solver's completeness is obtained by exploiting
pf-domains and $\Pfun$/2 constraints. Further empirical assessment of the solver may
lead to review the current choices and provide additional inference rules in future
releases.

%%%%%%%%%%%%%%%%%%%%%%%%%%%%%%%%%%%%%%%%%%%%%%%%%%%%%%%%%%%%%%%%%%%%%%%%%%%%%
\section{\label{empirical}Empirical Assessment}
%%%%%%%%%%%%%%%%%%%%%%%%%%%%%%%%%%%%%%%%%%%%%%%%%%%%%%%%%%%%%%%%%%%%%%%%%%%%%

In this section we present how the new version of \setlog (i.e. 4.8.2-2) improves its
efficiency and effectiveness when solving formulas including partial functions and
their operators. To do so we have generated more than 2,000 goals, some of which
include partial functions and the related operators. These goals have been used to
evaluate \setlog 4.8.0 as a test case generator for \Fastest, a model-based testing
tool \cite{CristiaRossiSEFM13}. Besides, these goals have been generated by \Fastest from
10 different Z specifications, some of which are formalizations of real requirements
and, in general, they cover a wide range of applications---totalizing around 3,000
lines of Z code. These goals not only include partial functions, but also sets (in
particular intentional sets), integer and relational constraints. Thus, we consider
that they are a representative sample.

In this assessment, we want to know: $(i)$ how many satisfiable and unsatisfiable
goals are found by \setlog; $(ii)$ how long it takes to process all the goals;
$(iii)$ how \setlog performs in each task compared with version 4.8.0 (which do not
include partial functions as primitive constraints).

%The \LaTeX{} mark-up of these case studies along with the \Fastest scripts used during
%the experiments can be downloaded from
%\url{https://www.dropbox.com/s/txfj2ih84zlzass/experiments-setlog-ttf.tar.gz}.

%\subsection{Experiment Settings}

Experiments were run on a 4 core Intel Core\texttrademark{} i5-2410M CPU at 2.30GHz
with 4 Gb of main memory, running Linux Ubuntu 12.04 (Precise Pangolin) 32-bit with
kernel 3.2.0-80-generic-pae. \setlog 4.8.0 and 4.8.2-2 over SWI-Prolog 6.6.6 for i386
were used during the experiments. A 10 seconds timeout was set as the maximum time
that \setlog can spend to give an answer for a goal.

%\subsection{Empirical Results}

Table \ref{t:results} displays the results of the experiments. The meaning of the
columns is as follows: \tableheadline{Z Spec}, Z specification;
\tableheadline{Goals}, number of goals processed during the experiment;
\tableheadline{S}, number of satisfiable goals detected as satisfiable;
\tableheadline{U}, number of goals detected as unsatisfiable; \tableheadline{A},
percentage of goals for which \setlog{} gives a meaningful answer (i.e.
$\tableheadline{A} = 100 (\tableheadline{S} +
\tableheadline{U}) / \tableheadline{Goals}$); \tableheadline{T}, time spent by
\setlog{} during the entire execution.

\begin{table}
\caption{\label{t:results}Summary of empirical assessment}
\begin{minipage}{\textwidth}
\begin{tabular}{lrrrrrrrrr}
\hline\hline

%GFR \multirow{2}{*}{{\tableheadline{Z Spec}}} & \multirow{2}{*}{\tableheadline{Goals}} &

\tableheadline{Z Spec} & \tableheadline{Goals} &
\multicolumn{4}{c}{\tableheadline{4.8.0}} &
\multicolumn{4}{c}{\tableheadline{4.8.2-2}}
  \\\cmidrule(r){3-6} \cmidrule(l){7-10}
& & \multicolumn{1}{c}{\tableheadline{S}} & \multicolumn{1}{c}{\tableheadline{U}} &
\multicolumn{1}{c}{\tableheadline{A}} & \multicolumn{1}{c}{\tableheadline{T}} &
\multicolumn{1}{c}{\tableheadline{S}} & \multicolumn{1}{c}{\tableheadline{U}} &
\multicolumn{1}{c}{\tableheadline{A}} & \multicolumn{1}{c}{\tableheadline{T}}
\\\midrule
SWPDC & 196   & 97    & 26  & 63\% & 1,238 & 99    & 26 & 64\% & 1,402  \\
Plavis & 232   & 151    & 36 & 81\%  & 510 & 151    & 33 & 79\% &  510 \\
Scheduler & 205   & 27   & 85  & 55\%  & 945  & 38   & 161  & 97\%  & 125 \\
Sec. class & 36    & 20    & 16  & 100\%  & 11  & 20    & 14  & 94\%  & 31 \\
Bank (1) & 100   & 23    & 39  & 62\%  & 388  & 25    & 75  & 100\%  &  28 \\
Bank (3) & 104   & 50    & 35  & 82\%  & 211  & 52    & 49  & 97\%  &  64 \\
Lift & 17    & 17     & 0  & 100\%  & 6  & 17     & 0  & 100\%  &  6 \\
Launcher & 1,206 & 0 & 1,093  & 91\%  & 1,334  & 23 & 1,183  & 100\%  & 370  \\
Symb. table & 27    & 11    & 10  & 78\%  & 68  & 11    & 16  & 100\%  & 9 \\
Sensors & 16    & 7    & 3   & 63\%  & 54  & 8     & 8   & 100\%  & 5 \\
\midrule
\textbf{Totals} & 2,139 & 403 & 1,343 & -- & 4,769 & 444 & 1,565 & -- &  2,552 \\
\hline\hline
\end{tabular}
\end{minipage}
\end{table}

As can be seen, \setlog{} 4.8.2-2 outperforms 4.8.0 in almost all sets of goals.
In effect, in all sets but two (SWPDC and Sec. class) 4.8.2-2 gives more right
answers and in less time than 4.8.0. Note that 4.8.2-2 hits 100\% of right
answers in 5 sets of goals while 4.8.0 does it only in 2. Also note the
impressive time reduction in, for example, Launcher. Given that giving more
right answers in less time is the best behavior, we can define $QI$, for quality
index, as $ QI = \lfloor 100*\tableheadline{A} /\tableheadline{T} \rfloor$.
Then, the higher the $QI$ the better. \setlog 4.8.2-2 has higher or equal $QI$
than 4.8.0 in all but one set of goals.

In summary, the experimental results show that adding constraints for partial
functions as \setlog's primitive constraints greatly improves its efficiency
and effectiveness as a constraint solver for a very general theory of sets.

\section{Conclusions}\label{sec:conclusions}

In this paper we have shown how to integrate partial functions as first-class
citizens into the CLP language with sets \setlog. Since partial functions can be
viewed as sets, they are embedded quite smoothly into \setlog{}, and all facilities
for set manipulation offered by \setlog{} are immediately available to manipulate
partial functions as well. We have added to the language a very limited number of new
primitive constraints, specifically devoted to deal with partial functions and we have
provided sound and terminating rewriting procedures for them. The resulting
constraint solver either terminates with \false or with a disjunction of simplified
constraints which the solver cannot further simplify (i.e., irreducible
constraints). We have identified conditions under which the ability to generate such
a disjunction guarantees the satisfiability of the input constraint. Moreover, we
have defined a number of inference rules that allow the solver to detect, in many
cases, unsatisfiability even in the more general situations (e.g. without requiring
to specify an upper bound for the cardinality of partial functions).

For the future, there are two main correlated lines of work: $(i)$ identifying more
precisely the class of irreducible constraints which are guaranteed to be
satisfiable; so far this class is restricted to irreducible constraints not
containing pf-constraints, but it is likely to be enlarged to include pf-constraints
as well, at least of some specific form (e.g., those which contain only unbound
variables, thus excluding for instance the irreducible constraints of the form
$\Ran(X,\{\dots\})$) $(ii)$ defining new inference rules that allow further
``hidden'' properties of irreducible constraints to be made explicit, in order to
make constraint solving more and more ``precise''; that is, on the one hand, to allow
the solver to detect more and more unsatisfiable constraints and, on the other hand,
to allow the class of irreducible constraints whose satisfiability can be decided
without the need to perform any labeling operation to be enlarged as much as
possible.

%\bibliographystyle{acmtrans}
%\bibliography{/home/mcristia/escritos/biblio}
%\bibliography{AddingPartialFunctionsICLP2_REV1}

\end{document}